\documentclass[a4,12pt,fleqn]{article}
\usepackage{epsfig,times,latexsym,enumerate,lscape,flafter,amsmath,amssymb,graphicx}
\usepackage{multirow}
\usepackage{graphicx}
\begin{document}   

 \title{An Effective Information Retrieval for  Ambiguous Query}
\author{R.K. Roul\thanks{email: rkroul@bits-goa.ac.in} \hspace{0.1pt}  and S.K. Sahay\thanks{email: ssahay@bits-goa.ac.in}\\
{\small BITS, Pilani - K.K. Birla, Goa Campus, Zuarinagar, Goa - 403726, India.}}
\date{}
\maketitle

\begin{abstract}

 Search engine returns thousands of web pages for a single user query, in which most of them are not relevant. 
 In this context, effective information retrieval from the expanding web is a challenging task, in particular, if the query is ambiguous. The major question arises here is that how to get the relevant pages for an ambiguous query. We propose an approach for the effective result of an ambiguous query by forming community vector based on association concept of data minning using vector space model and the 
freedictionary. We develop clusters by computing the similarity between community vectors and document vectors formed from the extracted web pages by the search engine. We use Gensim package to implement the algorithm because of its simplicity and robust nature. Analysis shows that our approach is an effective way to form clusters for an ambiguous query.

\end{abstract}
\indent\indent\indent{\bf Keywords:} {\it Information Retrieval, Clustering, Vector space model, Gensim.}

\section{Introduction}

\par On the web, search engines are key for the information retrieval (IR) for any user query. However, resolving ambiguous query is a challenging task, hence a vibrant area of research. Due to short and ambiguity in the user query, retrieving the information as per the intention of user in large volume of web is not straight forward. The ambiguities in queries is due to the short query length, which is on an average is 2.33 times on a popular search engine [1]. In this context, Sanderson [2] reports that 7\%-23\% of the queries frequently occur in two search engines are ambiguous with the average length one. For e.g. the familiar word {Java} which is ambiguous as it has multiple senses viz. Java coffee, Java Island and Java programming language etc. In the user query, ambiguities can also exists which do not appear in surface. Because of such ambiguities, search engine generally does not understand in what context user is looking for the information. Hence, it returns huge amount of information, in which most of the retrieved pages are irrelevant to the user. These huge amount of heterogeneous information retrieve not only increases the burden for search engine but also decreases its performance.

\par In this paper we propose an approach to improve the effectiveness of search engine by making clusters of word sense based on association concept of data mining, using vector space model of Gensim [6] and the freedictionary [13]. The association concept on which the clusters has formed can be describe as follows.
Suppose, if user queried for the word {\bf Apple}, which is associated in multiple context viz. {\it computer, fruit, company etc}. Each of this context associated with {\bf Apple} is again associate with different word senses viz. computer is associated with the keyboard, mouse, monitor etc. Hence  computer can be taken as community vector or cluster whose components/elements are the associated words {\it keyboard, mouse, monitor, etc}. Here, each element in the cluster represent the sense of computer vector for apple. So, if a user looking for apple as a computer, s'he may look for `apple keyboard' or `apple mouse' or `apple monitor' etc. We use Minipar [16] to transform a complete sentence into a dependency tree and for the classification of words and phrases into lexical categories.

\par The paper is organized as follows. In section 2 we examine the related work on the information retrieval based on clustering technique. In section 3 we briefly discuss the Gensim package for the implementation of our approach. In section 4  we present our approach for the effective information retrieval in the context of user query. Section 5 contains analysis of the algorithm. Finally Section 6 is the conclusion of the paper.

\section{Related Work}
\par Ranking and Clustering are the two most popular methods for information retrieval on the web. In ranking, a model is designed using training data, such that model can sort new objects according to their relevance's. There are many ranking models [14] which can be roughly categorized as query-dependent and query-independent models. In the other method i.e. clustering, an unstructured set of objects form a group, based on the similarity among each other. One of the most popular algorithms on clustering is k-means algorithm. However, the problem of this algorithm is that an inappropriate choice of clusters (k) may yield poor results. In case of an ambiguous query, word sense discovery is one of the useful method for IR in which documents are clustered in corpus. Discovering word senses by clustering the words according to their distributional similarity is done by Patrick et al, 2002. The main drawback of this approach is that they require large training data to make proper cluster and its performance is based on cluster centroid, which changes whenever a new web page is added to it. Hence identifying relevant cluster will be a tedious work.

\par Herrera et al., 2010 gave an approach, which uses several features extracted from the document collection and query logs for automatically identifying the user’s goal behind their queries. This approach success to classifies the queries into different categories like navigational, informational and transactional (B. J. Jansen et al., 2008) but fails to classify the ambiguous query. As query logs has been used, it may raise privacy concerns as long sessions are recorded and may led to ethical issues surrounding the users data collections. Lilyaa et.al [15] uses statistical relational learning (SRL) for the short ambiguous query based only on a short glimpse of user search activity, captured in a brief search session.
Many research has been done to map user queries to a set of categories (Powell et al., 2003; Dolin et al., 1998; Yu et al., 2001).
But all of the above techniques fails to identify the user intention behind the user query.

 \par The Word Sense Induction (Roberto Navigli et.al, 2010) method is a graph based clustering algorithm, in which snippets are clustered based on dynamic and finer grained notion of sense. The approach (Ahmed Sameh et al, 2010) with the help of modified Lingo algorithm, identifying frequent phrases as a candidate cluster label, the snippets are assigned to those labels. In this approach semantic recognition is identified by WordNet which enables recognition of synonyms in snippets. Clusters formation by the above two approaches not contain all the relevant pages of user choice. Our work uses free dictionary and association concept of data mining has been added to our approach to form clusters. Secondly it can handle the dynamic nature of the web as Gensim has been used. Hence the user intention behind the ambiguous query can be identified in simple and efficient manner.

\par In 2008, Jiyang Chen et. al. purposed an unsupervised approach to cluster results by word sense communities.  Clusters are made based on dependency based keywords which are extracted for large corpus and manual label are assigned to each cluster. In this paper we form the community vector and eliminate the problem of manual assignment of the cluster lable. We use Gensim package to avoid the dependency of the large training corpus size [5], and its ease of implementing vector space model (e.g. LSI, LDA). 

\section{Gensim} 

\par Gensim package is a python library for vector space modeling,  aims to process raw, unstructured digital texts (``plain text"). It can automatically extract semantic topics from documents, used basically for the Natural Language Processing (NLP) community. Its memory (RAM) independent feature with respect to the corpus size allows to process large web based corpora. In Gensism one can easily plugin his own input corpus and data stream and other vector space algorithms can be  trivially incorporated in it.

\par In Gensim, many unsupervised algorithms are based on word co-occurrence patterns within a corpus of training documents. Once these statistical patterns are found, any plain text documents can be succinctly expressed in the new semantic representation and can be queried for the topical similarity against other documents and so on. In addition it has following salient features
\begin{itemize}
\item{Straightforward interfaces, scalable software framework, low API learning curve and prototyping.}

\item{Efficient implementations of several popular vector space algorithms, calculation of TF-IDF (term frequency-inverse document frequency), distributed incremental Latent Semantic Analysis, distributed incremental incremental Latent Dirichlet Allocation(LDA).}
\item{I/O wrappers and converters around several popular data formats.}
\end{itemize}
\clearpage

\noindent{\bf Vector Space Model:}

\par In vector space model, each document is defined as a multidimensional vector of keywords in euclidean space whose axis correspond to the keyword i.e., each dimension corresponds to a separate keyword [4]. The keywords are extracted from the document and weight associated with each keyword determines the importance of the keyword in the document. Thus, a document is represented as, \\

$$ D_j = (w_{1j}, w_{2j}, w_{3j}, w_{4j}, ..........w_{nj})$$

where $w_{ij}$ is the weight of term $i$ in document $j$ indicating the relevance and importance of the keyword.\\

\noindent{\it TF-IDF Concept:}  TF is the measure of how often a word appears in a document and IDF is the measure of the rarity of a word within the search index. 
Combining TF-IDF is used to measure the statistical strength of the given word in reference to the query. Mathematically,                       
                       $$   {\text{TF}_i} = \frac{n_i}{\sum_k n_k}$$
 where,  $n_i$ is the number of occurrences of the considered terms and $n_k$ is the number of occurrences of all terms in the given document

                    $${\text{IDF}_i}= \log\frac{N}{df_i}$$
where, $N$ is  the number of occurrences of the considered terms and $df_i$ is the number of documents that contain term $i$.
 
                   $${\text{TF-IDF}} = {\text{TF}}_i \times \log\frac{N}{df_i}$$

\noindent{\it Cosine Similarity Measure:}  It is a technique to measure the similarity between the document and the query. The angle ($\theta$) between the document vector and the query vector determines the similarity between the document and the query and it is written as
\begin{equation}
\cos\theta = \frac{\sum w_{q,j}w_{ij}} {\sqrt{\sum w^2_{q,j}} \sqrt {\sum w^2_{i,j}}}
\end{equation}
\noindent $\sqrt{\sum w^2_{q,j}}$ and $\sqrt {\sum w^2_{i,j}}$ is the length of the query and document vector respectively.\\

\par If $\theta  = 0^\circ$ then the document and query is similar. As $\theta$ changes from $0^o$ to $90^o$, the similarity between the document and query decreases i.e. ${\mathbf D_2}$ will be more similar to query than ${\mathbf D_1}$, if the angle between ${\mathbf D_2}$ and query is smaller than the angel between ${\mathbf D_1}$ and query.

\section{Our Approach }

Our approach for an ambiguous query is described below in five steps and depicted in the flow chart (Fig. 1).

\begin{enumerate}
\item {\it Web page extraction and preprocessing:} Submit the ambiguous query to a search engine and extract top $\mathbf{n}$ pages. Preprocess the retrieve corpus as follows:
\begin{itemize}
\item Remove the stop and unwanted words.
\item Select noun as the keywords from the corpus using Minipar [16] and ignore other categories, such as verbs, adjectives, adverbs and pronounce. 
\item Do stemming using porter algorithm [12].
\item Save each processed $\mathbf{n}$ pages as documents $\mathbf{D_k}$, where $k = 1,2,3,.....n$.
\end{itemize}

\item  {\it Document vectors:} Compute TF and IDF score for all the keywords of each $\mathbf{D_k}$ using Gensim and make document vectors  of all the retrieved pages.

\item {\it Cluster formation:} We use the freedictionary with the option {\it\bf start with} to form the community vector of the queried word as follows
\begin{itemize}
\item Submit the ambiguous query (say apple) to the freedictionary, preprocess the retrieved data i.e. remove the queried, stop \& unwanted words. After stemming, save all the noun as keywords ($\mathbf{W_j}$) in a file $\mathbf{F_c}$, where $j = 1,2,3, ......m$

\item Now submit each $\mathbf{W_j}$ again to the freedictionary, preprocess the retrieved data and save the noun as keywords along with the queried word in a community file $F_\mathbf{W_j}$.

\item Search all the words of $F_\mathbf{W_j}$ in $\mathbf{D_k}$ using regular expression search technique. 

\item Delete those words in $F_\mathbf{W_j}$ which are not present in $\mathbf{D_k}$.

\item $\mathbf{W_j}$ is the formed community vectors (clusters) whose elements are the words saved in the file $F_\mathbf{W_j}$

\item Compute TF-IDF for each word in $F_\mathbf{W_j}$ in compare with $\mathbf{D_k}$to form community vectors.

\end{itemize}
 
\item { \it Similarity check:} Compute the cosine similarities between the formed documents and community vectors using eq. 1.

\item {\it Assignment of Documents to the Clusters:} Assign the documents to that cluster which has maximum similarity.
\end{enumerate}

\begin{figure}[!t]
   \centering
   \includegraphics[width=6.150in]{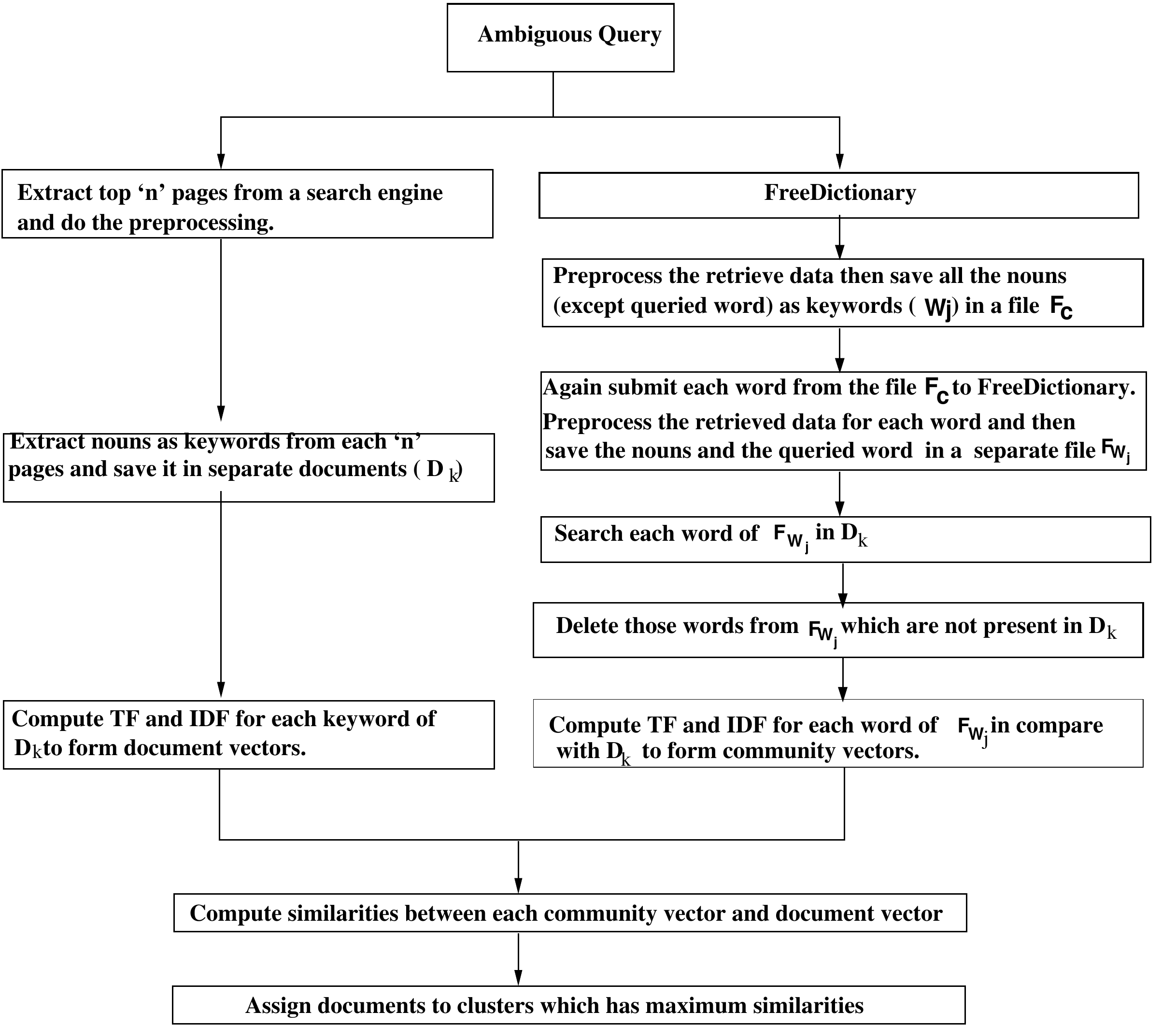}
 \caption{An effective IR for an ambiguous query}
\end{figure}

\section{Test Results}

\par To illustrate our approach we took four sample documents as shown in Table 1. We preprocess the documents and extracted ten keywords (apple,  computer, tree, keyboard, mouse, juice, country, vegetables, fruit, monitor) from the sample (Table 2). After assigning a token ID  to each selected keyword (Table 3)  TF \& IDF are computed which is shown in Table 4. In Table 5 computed  weight (TF-IDF) of all the four sample documents are given.With the calculated weight and respective token IDs, document vectors are generated (Table 6). 

\par The community vectors are formed as described in the section 4 (Table 7) and the corresponding TF-IDF and weights are calculated (Table 8). Cosine similarity are calculated defined by the eq. 1. Now the similarity between each community vector (C1, C2) and the set of document vectors (D1, D2, D3 and D4) are computed and maximum values of the similarity between community and document vectors form the cluster. From our experimental result, we found that (D1, D3 ) and (D2, D4)  associated with C1 and C2 respectively i.e. two clusters are generated (Table 9 and 10).

\par As an example, from the Table 10 we say that if the user search the ambiguous word apple, s'he will get two clusters C1 and C2, containing most relevant documents.

\section{Conclusion}

For an ambiguous query, we propose an effective approach for the IR by forming the clusters of relevant web pages. For cluster formation we use standard vector space model and the freedictionary. From our approach we find that user intention behind ambiguous query can be identify significantly. This unsupervised approach not only handles the corpus by extracting and analyzing significant terms, but also form desire clusters for real time query. Further we would extend our work for the multi word query and improving these clusters using ranking techniques.

\section*{Acknowledgment}

We are thankful to  Bharat Deshpande and our colleague Aruna Govada and K.V. Santhilata for their useful discussions and valuable suggestions.

\appendix

\section*{Appendix}
\begin{table}[hbt]
\label{table1}
\centering
{\begin{tabular}{|c|l|}
\hline
D1 &apple computer released new wireless keyboard and apple trees\\
&  are more in our country. \\
\hline
 D2 & all vegetables trees are different from apple trees.  \\
\hline
D3 & the apple mouse is a multi-button USB mouse manufactured and \\
&sold by apple Inc.\\
\hline
D4 & apple as juice or fruit is very tasty and apple launch new LED monitor.\\
\hline
\end{tabular}}
\caption{Sample documents taken for experiment.}
\end{table}
\begin{table}[h]
\label{table2}
\centering
{\begin{tabular}{|c|l|}
\hline
D1 & apple computer keyboard apple tree country\\
\hline
 D2 & vegetable tree apple tree \\
\hline
D3 & apple mouse mouse apple\\
\hline
D4 & apple juice fruit apple monitor \\
\hline
\end{tabular}}
\caption{Documents after preprocessing.}
\end{table}
\clearpage
\begin{table}[hbt]
\label{table3}
\centering
{\begin{tabular}{|l|c|}
\hline
\bf{Keyword} &\bf{Token ID}\\
\hline
apple & 0\\
\hline
computer & 1\\
\hline
tree &2\\
\hline
keyboard & 3\\
\hline
mouse & 4\\\hline
juice &5\\
\hline
country & 6\\
\hline
vegetable &7\\
\hline
fruit &8 \\
\hline
monitor & 9 \\
\hline
\end{tabular}}
\caption{Keywords \& respective token IDs.}
\end{table}

\begin{table}[hbt]
\label{table4}
\centering
{\begin{tabular}{|l|c|l|l|l|l|l|l|l|l|}
\hline
Keyword & D1 & TF1 & D2 & TF2 & D3 & TF3 & D4 & TF4 & IDF \\
\hline
apple & 2 & 0.33 & 1 & 0.25 & 2 & 0.5 & 2 & 0.4 & 0  \\
\hline
computer & 1 & 0.16 & 0 & 0 & 0 & 0 & 0 & 0 & 0.602 \\
\hline
tree & 1 & 0.16 & 2 & 0.5 & 0 & 0 & 0 & 0 &0.301\\
\hline
keyboard & 1 & 0.16 & 0 & 0 & 0 & 0 & 0 & 0 & 0.602 \\
\hline
mouse & 0 & 0 & 0 & 0 & 2 & 0.5 & 0 & 0 & 0.602 \\
\hline 
juice & 0 & 0 & 0 & 0 & 0 & 0 & 1 & 0.2 &0.602 \\
\hline
country & 1 & 0.16 & 0 & 0 & 0 & 0 & 0 & 0 & 0.602\\
\hline
vegetable & 0 & 0 & 1 & 0.25 & 0 & 0 & 0 & 0 & 0.602\\
\hline
fruit & 0 & 0 & 0 & 0 & 0 & 0 & 1 & 0.2 & 0.602 \\
\hline
monitor & 0 & 0 & 0 & 0 & 0 & 0 & 1 & 0.2 & 0.602 \\
\hline
\end{tabular}}
\caption{Calculation of TF-IDF for each documents.}
\end{table}
\clearpage
\begin{table}[hbt]
\label{table6a}
\centering
{\begin{tabular}{|l|l|l|l|l|}
\hline
Keyword & D1 & D2 & D3 & D4 \\
\hline
apple & 0 & 0 & 0 & 0\\
\hline
computer & 0.09632 & 0 & 0  & 0 \\
\hline
tree & 0.04816 & 0.1505 & 0 & 0 \\
\hline
keyboard & 0.09632 & 0 & 0 & 0 \\
\hline
mouse & 0 & 0 & 0.301 & 0 \\
\hline
juice & 0 & 0  &0 & 0.1204 \\
\hline
country & 0.09632 & 0 & 0 & 0 \\
\hline
vegetable & 0 & 0.1505 & 0 & 0 \\
\hline
fruit & 0 & 0 & 0 & 0.1204 \\
\hline
monitor & 0 & 0 & 0 & 0.1204\\
\hline
\end{tabular}}
\caption{Weight: TF x IDF.}
\end{table}
\begin{table}[hbt]
\label{table5}
\centering
{\begin{tabular}{|l|l|}
\hline
Documents & Corresponding Document Vectors \\
\hline 
D1 &
[(0, 0), (1, 0.09632), (2, 0.04816), (3, 0.09632), (4, 0), \\
& (5, 0), (6, 0.09632), (7, 0), (8, 0), (9, 0)]\\
\hline
D2 & [(0, 0), (1, 0), (2, 0.1505), (3, 0), (4, 0), (5, 0),  \\
&(6, 0), (7, 0.1505), (8, 0), (9, 0)]\\
\hline
D3 & [(0, 0), (1, 0), (2, 0), (3, 0), (4, 0.301), (5, 0), (6, 0),  \\
&(7, 0), (8, 0), (9, 0)] \\
\hline
D4 & [(0, 0), (1, 0), (2, 0), (3, 0), (4, 0), (5, 0.1204), (6, 0), \\
&(7, 0), (8, 0.1204), (9, 0.1204)]\\
\hline
\end{tabular}}
\caption{Representation of document as vectors.}
\end{table}
\begin{table}[hbt]
\label{table6}
\centering
{\begin{tabular}{|l|l|l|}
\hline
Community Vector  & Associated Keywords & [(ID,Frequency)]\\
\hline
Computer (C1) & computer, keyboard, & [(1,1), (3,1), (4,2), (9,1)]\\
&mouse, monitor &\\
 \hline
Fruit (C2) & fruit, tree, vegetable, juice & [(8,1), (2,3), (7,1), (5,1)]\\
\hline
\end{tabular}}
\caption{Community vectors formed from communities as [ID, Frequency].}
\end{table}

\begin{table}[hbt]
\label{table7}
\centering
{\begin{tabular}{|l|l|l|l|l|l|l|l|}
\hline
Keyword & C1 & TF C1 & C2 & TF C2 & IDF &\multicolumn{2}{c|}{Weight = TF$\times$ IDF}\\ 
\cline{7-8}
&&&&&& C1 & C2 \\
\hline
apple & 0 & 0 & 0 & 0 & 0 & 0 & 0\\
\hline
computer & 1 & 0.25 & 0 & 0 & 0.602 & 0.1505 & 0\\
\hline
tree & 0 & 0 & 3 & 0.75 &  0.301 & 0 & 0.22575 \\
\hline
keyboard & 1 & 0.25 & 0 & 0 & 0.602 & 0.1505 & 0\\
\hline
mouse & 2 & 0.5 & 0 & 0 & 0.602 & 0.301 & 0\\
\hline
juice & 0 & 0 & 1 & 0.25 & 0.602 & 0 & 0.1505 \\
\hline
country & 0 & 0 & 0 & 0 & 0.602 & 0 & 0 \\
\hline
vegetable & 0 & 0 & 1 & 0.25 & 0.602 &0 & 0.1505 \\
\hline
fruit & 0 & 0 & 1 & 0.25 & 0.602 & 0 & 0.1505 \\
\hline
monitor & 1 & 0.25 & 0 & 0 & 0.602 & 0.1505 &0 \\
\hline
\end{tabular}}
\caption{TF-IDF calculation for community vector.}
\end{table}

\begin{table}[hbt]
\label{table8}
\centering
{\begin{tabular}{|l|l|l|l|}
\hline
Document/Community & C1 & C2 & Resultant Cluster \\
&&&(Max(C1,C2)) \\
\hline
D1 & 0.41939 & 0.18165 & C1 (Computer) \\
\hline 
D2 & 0.0 & 0.77149 & C2 (Fruit) \\
\hline 
D3 & 0.75593 & 0.0 & C1 (Computer) \\
\hline 
D4 & 0.21828 & 0.5041 & C2 (Fruit)\\
\hline
\end{tabular}}
\caption{Similarity between each community and document is tabulated.}
\end{table}
\begin{table}[!h]
\label{table9}
\centering
{\begin{tabular}{|l|l|l|}
\hline
Query (apple) & Community (sense) & Cluster\\
\hline
Cluster 1  & Computer & D1, D3\\
\hline
Cluster 2 & Fruit & D2, D4\\
\hline
\end{tabular}}
\caption{Final clustering of relevant documents.}
\end{table}

\end{document}